\documentclass[12pt]{article}
\usepackage{graphicx}
\usepackage{epsfig}
\usepackage{latexsym}
\usepackage{latexsym}
\usepackage{amsmath}

\begin{document}

\author{Khireddine Nouicer\thanks{%
e-mail:khnouicer@mail.univ-jijel.dz} \\
Department of Physics, Laboratory of Theoretical Physics (L.P.Th),\\
Faculty of Sciences, University of Jijel,\\
Bp 98, Ouled Aissa, 18000 Jijel, Algeria. }
\title{Effect of Minimal lengths on Electron Magnetism}
\maketitle

\begin{abstract}
We study the magnetic properties of electron in a constant magnetic field
and confined by a isotropic two dimensional harmonic oscillator on a space
where the coordinates and momenta operators obey generalized commutation
relations leading to the appearance of a minimal length. Using the momentum
space representation we determine exactly the energy eigenvalues and
eigenfunctions. We prove that the usual degeneracy of Landau levels is
removed by the presence of the minimal length in the limits of weak and
strong magnetic field.The thermodynamical properties of the system, at high
temperature, are also investigated showing a new magnetic behavior in terms
of the minimal length.
\end{abstract}


\bigskip

\section{Introduction}

In a series of papers Kempf et al. \cite{kempf0, kempf1, kempf2,kempf3}
introduced a deformed quantum mechanics based on modified commutation
relation between position and momentum operators. These commutation relation
lead to generalized Heisenberg uncertainty principle (GUP) which define non
zero minimum length in position or minimal length. The concepts of GUP and
minimal length originate from several studies in string theory \cite{groos},
loop quantum gravity \cite{garay} and non-commutative field theories \cite%
{nekrasov}. Other similar constructions leading to the concept of GUP have
been also initiated by some authors \cite{hos01,hos02,hos03,sastry}. The
fundamental outcome of the GUP is the appearance of an UV/IR "bootstrap".
This mixing between UV and IR divergences, first noticed in the AdS/CFT
correspondence \cite{suskind}, is also a feature of non-commutative quantum
field theory \cite{micu}. On the other some scenarios have been proposed
where the minimal length is related to large extra dimensions \cite{hos01},
to the running coupling constant \cite{hos02} and to the physics of black
holes production \cite{hos03}.

Recently, a great interest has been devoted to studies of quantum systems in
the presence of minimal lengths. The solution of Schrodinger equation in
momentum space for the harmonic oscillator in D-dimensions \cite%
{kempf1,kempf2,minic} and the cosmological constant problem with minimal
lengths have been investigated in \cite{chang,chang01}. Furthermore, the
effect of the minimal length on the energy spectrum and momentum wave
functions of the Coulomb potential in one dimension and three dimensions has
been studied respectively in \cite{fytio} and \cite{brau,akhoury,Benzik},
the high temperature properties of the one dimensional Dirac oscillator has
been investigated by the author in \cite{nouicer0}, the solution of the
three dimensional Dirac oscillator using supersymmetric quantum mechanics
\cite{quesne} and the Casimir force for the electromagnetic field in the
presence of the minimal length has been also computed \cite{nouicer1,hossen}.

In this paper we are interested by the effect of the minimal length on the
electron magnetism confined by a harmonic potential in the regime of high
temperatures. The electron magnetism under confining potential has been
considered in the ordinary case in \cite{ishi, Gazeau} and recently in the
context of canonical non-commutative quantum mechanics in \cite{dayi,Jellal}%
. The rest of the paper is organized as follow. In section II, we give a
brief review of quantum mechanics with generalized commutation relations and
solve exactly the stationary Schrodinger equation, in the momentum space
representation, for a spinless electron under the action of a constant
magnetic field and a isotropic harmonic oscillator. In section III, the
magnetic moment and susceptibility of the system are examined in the regime
of high temperature. Section V is left for concluding remarks.

\section{Electron in the presence of minimal lengths}

Let us start with the following generalized D-dimensional commutation
relations \cite{kempf1}

\begin{eqnarray}
\left[ X_{i},P_{j}\right] &=&i\hbar \left( \delta _{ij}+\delta _{ij}\beta
P^{2}+\beta ^{\prime }P_{i}P_{j}\right) ,  \label{1} \\
\left[ X_{i},X_{j}\right] &=&-i\hbar \left[ \left( 2\beta -\beta ^{\prime
}\right) +\left( 2\beta +\beta ^{\prime }\right) \beta P^{2}\right] \epsilon
_{ijk}L_{k}, \\
\left[ P_{i},P_{j}\right] &=&0,
\end{eqnarray}%
with $\beta $ and $\beta ^{\prime }$ two very small non negative parameters
and $D$ the space dimension. The components of the angular momentum given by
\begin{equation}
L_{i}=\frac{1}{1+\beta P^{2}}\epsilon _{ijk}X_{j}P_{k},  \label{ang}
\end{equation}%
satisfy the usual commutation relations
\begin{equation}
\left[ L_{i},X_{j}\right] =i\hbar \epsilon _{ijk}X_{k},\qquad \left[
L_{i},P_{j}\right] =i\hbar \epsilon _{ijk}P_{k}.
\end{equation}%
Using the fact that $<P_{i}>=0$, and that $\left( \Delta P_{i}\right) $ is
isotropic we easily obtain the generalized uncertainty principle (GUP)
\begin{equation}
\left( \Delta X_{i}\right) \left( \Delta P_{i}\right) \geq \frac{\hbar }{2}%
\left( 1+\beta D\left( \Delta P_{i}\right) ^{2}+\beta ^{\prime }\left(
\Delta P_{i}\right) ^{2}\right) .
\end{equation}%
A minimization of the saturate GUP with respect to $\Delta P_{i}$ gives an
isotropic minimal length
\begin{equation}
\left( \Delta X_{i}\right) _{min}=\hbar \sqrt{D\beta +\beta ^{\prime }}%
,\qquad i=1,2,3,\cdots D.
\end{equation}%
This relation implies a lost of the notion of localization in the position
space. Since we are going to work in momentum space we use the following
representation of the position and momentum operator%
\begin{equation}
X_{i}=i\hbar \left[ \left( 1+\beta p^{2}\right) \frac{\partial }{\partial
p_{i}}+\beta ^{\prime }p_{i}p_{j}\frac{\partial }{\partial p_{j}}+\gamma
p_{i}\right] ,\quad P_{i}=p_{i},\quad L_{i}=-i\hbar \varepsilon _{ijk}p_{j}%
\frac{\partial }{\partial p_{k}}.  \label{x-p}
\end{equation}

The parameter $\gamma $ does not affect the commutation relations and only
modify the squeezing factor of the momentum space measure. In fact the inner
product is now defined as
\begin{equation}
\int \frac{d^{D}p}{\left( 1+\left( \beta +\beta ^{\prime }\right)
p^{2}\right) ^{1-\alpha }}\mid p><p\mid ,\qquad \alpha =\frac{\gamma -\beta
^{\prime }\frac{D-1}{2}}{\beta +\beta ^{\prime }},
\end{equation}
In the following we use the simple algebra with $\beta ^{\prime }=0$ and $%
\gamma =0$.

Let us consider a spinless electron under the action of a constant magnetic
field and confined by a two dimensional isotropic harmonic oscillator of
frequency $\omega _{0}$. This system is described by the following
Hamiltonian

\begin{equation}
H=\frac{1}{2m}\left( \mathbf{P}-\frac{q}{c}\mathbf{A}\right) ^{2}+\frac{%
m\omega _{0}^{2}}{2}(X^{2}+Y^{2}),  \label{H1}
\end{equation}%
where $\mathbf{A}$ is the vector potential. In the symmetric gauge $\mathbf{A%
}$ is given by
\begin{equation}
\mathbf{A}=\frac{B_{0}}{2}\left( -y\mathbf{i}+x\mathbf{j}\right) ,
\end{equation}%
where $B_{0}$ is the magnitude of the magnetic field.

Using the commutation relation $\left( \ref{1}\right) ,$ the Hamiltonian is
then written as
\begin{equation}
H=\frac{P_{z}^{2}}{2m}+\frac{P_{x}^{2}+P_{y}^{2}}{2m}+\frac{m\tilde{\omega}%
^{2}}{2}(X^{2}+Y^{2})+\frac{\omega }{2}L_{z},  \label{H2}
\end{equation}%
with $\tilde{\omega}=\sqrt{\omega ^{2}+\omega _{0}^{2}}$ , $\omega =\frac{%
qB_{0}}{2mc}$ the cyclotron frequency and $L_{z}=-i\hbar \left(
p_{x}\partial _{p_{y}}-p_{y}\partial _{p_{x}}\right) .$

In the appendix the solutions to the eigenvalue equation $H\Psi _{nl }\left(
\mathbf{p}\right) =E_{nl}\Psi _{ nl}\left( \mathbf{p}\right) $ are found and
the radial momentum wave functions given by

\begin{equation}
R_{nl}(p)={\mathcal{N}}(1+\beta p^{2})^{-\frac{\lambda +\left\vert
l\right\vert }{2}}(\beta p^{2})^{\frac{\left\vert l\right\vert }{2}%
}P_{n}^{(\lambda -1,\left\vert l\right\vert )}\left( \frac{\beta p^{2}-1}{%
\beta p^{2}+1}\right) ,
\end{equation}%
with $\lambda $ given by $\left( \ref{lam}\right) $. The constant $\mathcal{N%
}$ is calculated by employing the normalization condition $\int \frac{d^{3}p%
}{(1+\beta p^{2})}\mid R_{nl}(p)\mid ^{2}=1$ and the Jacobi polynomials
orthogonality relation \cite{grad}. Finally the normalized radial momentum
wave functions are given by%
\begin{equation}
R_{nl}(p)=\sqrt{\beta }\sqrt{\frac{2(n!)(2n+\lambda +\left\vert l\right\vert
)\Gamma (n+\lambda +\left\vert l\right\vert )}{\Gamma (n+\lambda )\Gamma
(n+\left\vert l\right\vert +1)}}(1+\beta p^{2})^{-\frac{\lambda +\left\vert
l\right\vert }{2}}(\beta p^{2})^{\frac{\left\vert l\right\vert }{2}%
}P_{n}^{(\lambda -1,\left\vert l\right\vert )}\left( \frac{\beta p^{2}-1}{%
\beta p^{2}+1}\right) .  \label{wave}
\end{equation}%
However, as pointed in \cite{kempf1}, the normalization condition alone does
not guarantied physically relevant wavefunctions but the latter must be in
the domain of $p$, which physically means that it should have a finite
uncertainty in momentum. This leads to the condition

\begin{equation}
\langle p^{2}\rangle =\int_{0}^{\infty }\frac{p^3dp}{1+\beta p^{2}}%
\Bigl|R_{nl}(p)\Bigr|^{2}<\infty .  \label{cond}
\end{equation}%
In our case the integrand in (\ref{cond}) behaves like $p^{-2\lambda +1}$
which requires $\lambda >\frac{1}{2}$. Then we choose the upper sign in the
expression of $\lambda $. However the condition $\lambda >\frac{1}{2}$ can
be also obtained from physical considerations. Let us take $\lambda $ with
the minus and work with $l=0$
\begin{equation}
\lambda =1-\frac{1}{m\tilde{\omega}\hbar \beta }.
\end{equation}%
Using the fact that $\hbar \sqrt{2\beta }\leq l_{c}$ where $l_{c}=\sqrt{%
\frac{2\hbar }{m\tilde{\omega}}}$ is the characteristic length of the
oscillator, we get
\begin{equation}
m\hbar \tilde{\omega}\beta =\frac{\left( \Delta X\right) _{\text{min}}^{2}}{%
l_{\text{c}}^{2}}<1,
\end{equation}%
and then the condition $\lambda >\frac{1}{2}$ is not satisfied.

The energy spectrum is now derived from equation (\ref{cond0})
\begin{equation}
\frac{\mathcal{E}_{nl}-\Omega }{\beta }=2(N+1)(\lambda
-1)+(N^{2}+l^{2}+2N+2),
\end{equation}%
where $N=2n+\left\vert l\right\vert $ is the principal quantum number. Using
the expressions of $\kappa $, $\mathcal{E}_{nl}$ and $\Omega $ we finally
obtain
\begin{equation}
E_{nl}=\frac{p_{z}^{2}}{2m}+\hbar \tilde{\omega}\left[ (N+1)\sqrt{1+\left(
m\hbar \tilde{\omega}\beta \right) ^{2}\left( 1+l^{2}\right) }+\frac{m\hbar
\tilde{\omega}\beta }{2}(N^{2}+l^{2}+2N+2)+\frac{\omega }{2\tilde{\omega}}l%
\right].
\end{equation}%
Ignoring the contribution of $\frac{p_{z}^{2}}{2m}$ and setting $\omega
=\omega _{0}$ we reproduce exactly the energy spectrum of the two
dimensional harmonic oscillator with minimal length \cite{minic}. A Taylor
expansion to first order in $m\hbar \tilde{\omega}\beta $ gives
\begin{equation}
E_{nl}=\frac{p_{z}^{2}}{2m}+\hbar \tilde{\omega}\left[ (N+1)+\frac{\beta
m\hbar \tilde{\omega}}{2}(N^{2}+l^{2}+2N+2)+\frac{\omega l}{2\tilde{\omega}}%
\right] .
\end{equation}%
Introducing the following quantum numbers
\begin{equation}
n_{d}=n+\frac{\left\vert l\right\vert +l}{2},\quad n_{g}=n+\frac{\left\vert
l\right\vert -l}{2},
\end{equation}%
we obtain
\begin{eqnarray}
E_{n_{d},n_{g}} &=&\frac{p_{z}^{2}}{2m}+\hbar \tilde{\omega}\left( 1+\beta
m\hbar \tilde{\omega}\right) +\hbar \tilde{\omega}\left[ \left( 1+\beta
m\hbar \tilde{\omega}+\frac{\omega }{2\tilde{\omega}}\right) n_{d}+\beta
m\hbar \tilde{\omega}n_{d}^{2}\right]  \notag \\
&&+\hbar \tilde{\omega}\left[ \left( 1+\beta m\hbar \tilde{\omega}-\frac{%
\omega }{2\tilde{\omega}}\right) n_{g}+\beta m\hbar \tilde{\omega}n_{g}^{2}%
\right]  \label{energy}
\end{eqnarray}%
Let us in the following ignore the term $\frac{p_{z}^{2}}{2m}$ and examine
the degeneracy of Landau levels in some limiting cases of the magnetic field.

\begin{itemize}
\item \underline{\textit{Weak magnetic field}}.
\end{itemize}

This case corresponds to $\omega \ll \omega _{0}$ such that we have $\tilde{%
\omega}\approx \omega _{0}$. The energy spectrum is then approximated by
\begin{equation}
E_{\gamma ,\rho }\approx 2\hbar \omega _{0}\left[ (\gamma +\frac{1}{2}%
)+\beta m\hbar \omega _{0}\left[ \gamma \left( \gamma +1\right) +\rho ^{2}+%
\frac{1}{2}\right] \right] ,
\end{equation}%
where $\gamma =\frac{n_{d}+n_{g}}{2},\quad \rho =\frac{n_{d}-n_{g}}{2}.$ We
observe that the usual degeneracy of the Landau levels is removed by the
presence of the minimal length.

\begin{itemize}
\item \underline{\textit{Strong magnetic field.}}
\end{itemize}

In this case we have $\omega \gg \omega _{0}$, $\tilde{\omega}\approx \omega$
and the energy spectrum is given by%
\begin{equation}
E_{n_{d},n_{g}}\approx2\hbar \omega \left[ \left( \gamma +\frac{1+\rho }{2}%
\right) +\beta m\hbar \omega \left[ \gamma \left( \gamma +1\right) +\rho ^{2}%
\right] \right] .
\end{equation}%
In this case also we observe that the degeneracy is removed. This latter
result shows a difference with the commutative and the canonical
non-commutative cases respectively, where the degeneracy of the Landau
levels in the strong magnetic field limit is still present \cite%
{Gazeau,Jellal}.

\section{Thermodynamical properties}

In this section we are interested by the thermodynamical properties of the
system at high temperatures. We set $z=e^{\tilde{\beta}\mu }$ and $\tilde{%
\beta}=1/kT$, where $\mu $ is the chemical potential and use the following \
assumption $\tilde{\beta}$ $\left\vert \mu -\hbar \tilde{\omega}\right\vert
<<1.$

Let us start by computing the one particle states density $g(n_{d},n_{g})$
given by

\begin{equation}
g(n_{d},n_{g})=\frac{V^{\frac{2}{3}}}{4\pi \hbar ^{2}}\underset{E_{n}\leq
E<E_{n+1}}{\int }\frac{dp_{x}dp_{y}}{\left( 1+\beta p^{2}\right) },
\end{equation}%
with $V^{\frac{2}{3}}$ a surface term. In polar coordinates we obtain

\begin{equation}
g(n_{d},n_{g})=\frac{V^{\frac{2}{3}}}{2\pi \hbar ^{2}}\int_{p_{n}}^{p_{n+1}}%
\frac{pdp}{\left( 1+\beta p^{2}\right) }=\frac{V^{\frac{2}{3}}}{4\pi \beta
\hbar ^{2}}\ln \frac{1+\beta p^{2}(n_{d}+1,n_{g}+1)}{1+\beta
p^{2}(n_{d},n_{g})}.
\end{equation}%
Using

\begin{equation*}
p(n_{d},n_{g})=\sqrt{2m\hbar \tilde{\omega}}\left[ \left( 1+\beta m\hbar
\tilde{\omega}\right) \left( n_{d}+n_{g}\right) +\frac{\omega }{2\tilde{%
\omega}}\left( n_{d}-n_{g}\right) +\beta m\hbar \tilde{\omega}\left(
n_{d}^{2}+n_{g}^{2}\right) \right] ^{\frac{1}{2}}
\end{equation*}%
and the fact that $1+\beta m\hbar \tilde{\omega}\approx 1,$ by virtue of the
GUP, we obtain

\begin{equation}
g(n_{d},n_{g})\approx g(\gamma )=\frac{V^{\frac{2}{3}}}{4\pi \beta \hbar ^{2}%
}\ln \left[ 1+4m\beta \hbar \tilde{\omega}\left[ 1+\beta m\hbar \tilde{\omega%
}\left( 2\gamma +1\right) \right] \right] .  \label{g}
\end{equation}
The expression of the one particle states density in the standard situation
is obtained by taking the limit $\beta \rightarrow 0$

\begin{equation}
g=\frac{m\tilde{\omega}V^{\frac{2}{3}}}{\pi \hbar }.
\end{equation}%
The thermodynamical potential is defined by the following expression

\begin{equation}
\Phi =-\frac{V^{\frac{1}{3}}}{2\pi \tilde{\beta}\hbar }\int_{-\infty
}^{+\infty }\frac{dp_{z}}{1+\beta p_{z}^{2}}\sum_{\gamma =0}^{\infty
}g(\gamma )\ln \left[ 1+z\exp \left( -\tilde{\beta}E\right) \right] .
\end{equation}%
Using $g(\gamma )$ and $E$ \ given respectively by $\left( \ref{g}\right) $
and $\left( \ref{energy}\right) ,$ we obtain

\begin{eqnarray}
\Phi &\approx &-\frac{V}{8\pi ^{2}\tilde{\beta}\beta \hbar ^{3}}\exp \left( -%
\tilde{\beta}\hbar \tilde{\omega}\left( 1+\beta m\hbar \tilde{\omega}\right)
\right) \int_{-\infty }^{+\infty }\frac{dp_{z}}{1+\beta p_{z}^{2}}\exp
\left( -\tilde{\beta}\frac{p_{z}^{2}}{2m}\right)  \notag \\
&&\times \sum_{n_{d},n_{g}=0}^{\infty }\ln \left[ 1+4m\beta \hbar \tilde{%
\omega}\left[ 1+\beta m\hbar \tilde{\omega}\left( n_{d}+n_{g}+1\right) %
\right] \right]  \notag \\
&&\times \exp \left( -\tilde{\beta}\hbar \tilde{\omega}\left[ \left( 1+\beta
m\hbar \tilde{\omega}\right) \left( n_{d}+n_{g}\right) +\frac{\omega }{2%
\tilde{\omega}}\left( n_{d}-n_{g}\right) +\beta m\hbar \tilde{\omega}\left(
n_{d}^{2}+n_{g}^{2}\right) \right] \right) .
\end{eqnarray}%
Using the approximation $\ln \left( 1+ua\right) \approx ua$ \ we write $\Phi
$ as

\begin{eqnarray}
\Phi &\approx &-\frac{4m\beta \hbar \tilde{\omega}V}{8\pi ^{2}\tilde{\beta}%
\beta \hbar ^{3}}\exp \left( -\tilde{\beta}\hbar \tilde{\omega}\left(
1+\beta m\hbar \tilde{\omega}\right) \right) \int_{-\infty }^{+\infty }\frac{%
dp_{z}}{1+\beta p_{z}^{2}}\exp \left( -\tilde{\beta}\frac{p_{z}^{2}}{2m}%
\right)  \notag \\
&&\left[ \left( 1+\beta m\hbar \tilde{\omega}\right)
S_{1}^{+}S_{1}^{-}+\beta m\hbar \tilde{\omega}\left(
S_{2}^{+}S_{1}^{-}+S_{2}^{-}S_{1}^{+}\right) \right] ,
\end{eqnarray}%
with the sums $S_{1}^{\pm }$ and $S_{1}^{\pm }$ defined by

\begin{eqnarray}
S_{1}^{\pm } &=&\sum_{n}\left[ \exp \left( -\tilde{\beta}\hbar \tilde{\omega}%
\left[ \left( 1+\beta m\hbar \tilde{\omega}\pm \frac{\omega }{2\tilde{\omega}%
}\right) n+\beta m\hbar \tilde{\omega}n^{2}\right] \right) \right] , \\
S_{2}^{\pm } &=&\sum_{n}n\left[ \exp \left( -\tilde{\beta}\hbar \tilde{\omega%
}\left[ \left( 1+\beta m\hbar \tilde{\omega}\pm \frac{\omega }{2\tilde{\omega%
}}\right) n+\beta m\hbar \tilde{\omega}n^{2}\right] \right) \right] .
\end{eqnarray}%
These sums are computed using the Euler Formula given by

\begin{equation}
\sum\limits_{n=0}^{\infty }f(n)=\frac{f(0)}{2}+\int_{0}^{\infty
}f(x)dx-\sum_{p=1}^{\infty }\frac{1}{\left( 2p\right) !}B_{2p}f^{\left(
2p-1\right) }\left( 0\right) ,  \label{euler}
\end{equation}%
where $B_{2p}$ are Bernoulli's numbers and $f^{\left( 2p-1\right) }$ $\left(
0\right) $ derivatives of the function $f\left( x\right) $ at $x=0.$ In the
high temperatures regime the contribution of the third term in $\left( \ref%
{euler}\right) $ is negligible.

With the aid of the formula \cite{grad}
\begin{equation}
\int_{0}^{\infty }x^{\nu -1}e^{-px^{2}-qx}dx=\left( 2p\right) ^{-\frac{\nu }{%
2}}\Gamma \left( \nu \right) \exp \left( \frac{q^{2}}{8p}\right) D_{-\nu
}\left( \frac{q}{\sqrt{2p}}\right) ,
\end{equation}%
where $D_{\mu }\left( u\right) $ is the cylindrical function, performing the
integration over $p_{z}$ and using the approximation $1+\beta m\hbar \tilde{%
\omega}\approx 1$ we obtain
\begin{eqnarray}
\Phi  &\approx &-\frac{4m\tilde{\omega}V}{8\pi \sqrt{\beta }\tilde{\beta}%
\hbar ^{2}}\exp \left( -\tilde{\beta}\hbar \tilde{\omega}\right) \exp \left(
\frac{\tilde{\beta}}{2\beta m}\right) \left[ 1-\hbox{erf}\left( \sqrt{\frac{%
\tilde{\beta}}{2\beta m}}\right) \right]   \notag \\
&&\times \left[ \left( \frac{1}{2}+A_{1}^{+}\right) \left( \frac{1}{2}%
+A_{1}^{-}\right) +\beta m\hbar \tilde{\omega}\left( A_{2}^{+}\left( \frac{1%
}{2}+A_{1}^{-}\right) +A_{2}^{-}\left( \frac{1}{2}+A_{1}^{+}\right) \right) %
\right] ,  \label{thermo}
\end{eqnarray}%
where we have set
\begin{equation}
A_{1}^{\pm }=\frac{e^{\frac{u_{\pm }^{2}}{4}}}{\hbar \tilde{\omega}\sqrt{2%
\tilde{\beta}\beta m}}D_{-1}\left( u_{\pm }\right) ,\quad A_{2}^{\pm }=\frac{%
e^{\frac{u_{\pm }^{2}}{4}}}{2\hbar ^{2}\tilde{\omega}^{2}\tilde{\beta}\beta m%
}D_{-2}\left( u_{\pm }\right) ,\quad u_{\pm }=\frac{\left( 1\pm \frac{\omega
}{2\tilde{\omega}}\right) }{\sqrt{2}}\sqrt{\frac{\tilde{\beta}}{m\beta }}.
\end{equation}%
At this stage we make the physical assumption that $u_{\pm }$ is a large
parameter. In fact we rewrite $u_{\pm }$ as
\begin{equation}
u_{\pm }=\frac{\left( 1\pm \frac{\omega }{2\tilde{\omega}}\right) }{\sqrt{%
2\pi }}\frac{\lambda }{\left( \Delta X\right) _{\text{min}}},
\end{equation}%
where $\frac{\lambda }{\left( \Delta X\right) _{\text{min}}}$ is the ratio
between the thermal wave length $\lambda =\sqrt{\frac{2\pi \tilde{\beta}%
\hbar ^{2}}{m}}$ and the minimal length $\left( \Delta X\right) _{\text{min}%
}=\hbar \sqrt{2\beta }$ . The thermal wave length is a physical
characteristic length of the system and, in order to be experimentally
probed, must be larger than the minimal length. The latter assertion is the
physical content of the GUP and is expressed by

\begin{equation}
\frac{\lambda }{\left( \Delta X\right) _{\text{min}}}>1.  \label{cond2}
\end{equation}%
Then, for large values of $\frac{\tilde{\beta}}{m\beta }$ we use the
following approximations

\begin{equation}
e^{\frac{\tilde{\beta}}{2\beta m}}\left[ 1-\hbox{erf}\left( \sqrt{\frac{%
\tilde{\beta}}{2\beta m}}\right) \right] \approx \sqrt{\frac{2\beta m}{\pi
\tilde{\beta}}}\left[ 1-\frac{\beta m}{\tilde{\beta}}\right] ,
\end{equation}%
and

\begin{equation}
e^{\frac{u_{\pm }^{2}}{4}}D_{-1}\left( u_{\pm }\right) \approx \frac{\sqrt{%
2m\beta /\tilde{\beta}}}{\left( 1\pm \frac{\omega }{2\tilde{\omega}}\right) }%
,\quad e^{\frac{u_{\pm }^{2}}{4}}D_{-2}\left( u_{\pm }\right) \approx \frac{%
2m\beta /\tilde{\beta}}{\left( 1\pm \frac{\omega }{2\tilde{\omega}}\right)
^{2}},
\end{equation}%
which leads to%
\begin{equation}
A_{1}^{\pm }=\frac{1}{\hbar \tilde{\omega}\tilde{\beta}\left( 1\pm \frac{%
\omega }{2\tilde{\omega}}\right) },\quad A_{2}^{\pm }=\frac{1}{\hbar ^{2}%
\tilde{\omega}^{2}\tilde{\beta}^{2}\left( 1\pm \frac{\omega }{2\tilde{\omega}%
}\right) ^{2}}.
\end{equation}%
Using $\frac{1}{2}+A_{1}^{\pm }\approx A_{1}^{\pm }$, we finally obtain the
thermodynamical potential at high temperatures

\begin{equation}
\Phi \approx -\frac{2V}{\tilde{\beta}\hbar \lambda ^{3}}\left[ \left[ 1-%
\frac{m\beta }{\tilde{\beta}}\right] \frac{1}{\tilde{\omega}\left( 1-\left(
\frac{\omega }{2\tilde{\omega}}\right) ^{2}\right) }+\frac{2\beta m}{\tilde{%
\omega}\tilde{\beta}}\frac{1}{\left( 1-\left( \frac{\omega }{2\tilde{\omega}}%
\right) ^{2}\right) ^{2}}\right] .
\end{equation}%
The magnetic moment of the system defined by $M_{\beta }=-\frac{\partial }{%
\partial B}\Phi $ is then easily obtained%
\begin{equation}
M_{\beta }=-\frac{8V}{\tilde{\beta}\hbar \lambda ^{3}}\frac{q}{2mc}\frac{%
\omega }{\tilde{\omega}\left( 3\tilde{\omega}^{2}+\omega _{0}^{2}\right) ^{2}%
}\left[ \left( 3\tilde{\omega}^{2}-\omega _{0}^{2}\right) -\frac{m\beta }{%
\tilde{\beta}}\left[ \left( 3\tilde{\omega}^{2}-\omega _{0}^{2}\right) -%
\frac{24\tilde{\omega}^{2}\omega ^{2}}{\left( 3\tilde{\omega}^{2}+\omega
_{0}^{2}\right) }\right] \right] .  \label{m}
\end{equation}%
This expression is always negative otherwise we have $\beta <0.$ This case
is discarded since the parameter $\beta $ defines the minimal length which
is a physical scale. On the other hand the dependence of the magnetic moment
on the applied external magnetic field is not trivial and in order to
extract useful magnetic properties of the system we have shown, in figure 1,
the behavior of the susceptibility $\chi _{\beta }=\frac{\partial M}{%
\partial B}$ as a function of the magnetic field. We first observe that we
have two critical values $B_{1}$ and $B_{2}$ of the magnetic field for which
the effect of the minimal length is undetectable. For values smaller than $%
B_{1},$ corresponding to weak magnetic fields, the Landau diamagnetism is
less pronounced than in the standard situation without the minimal length.
In the strong magnetic field case corresponding to values larger than $B_{2},
$ the system exhibits a stronger paramagnetic behavior with increasing
values of the minimal length. A third regime, corresponding to intermediate
values of the magnetic field, the situation is inverted since the
diamagnetism and the paramagnetism behaviors are respectively stronger and
weaker with increasing values of the minimal length.

\begin{center}
\includegraphics[height=10cm,
width=10cm]{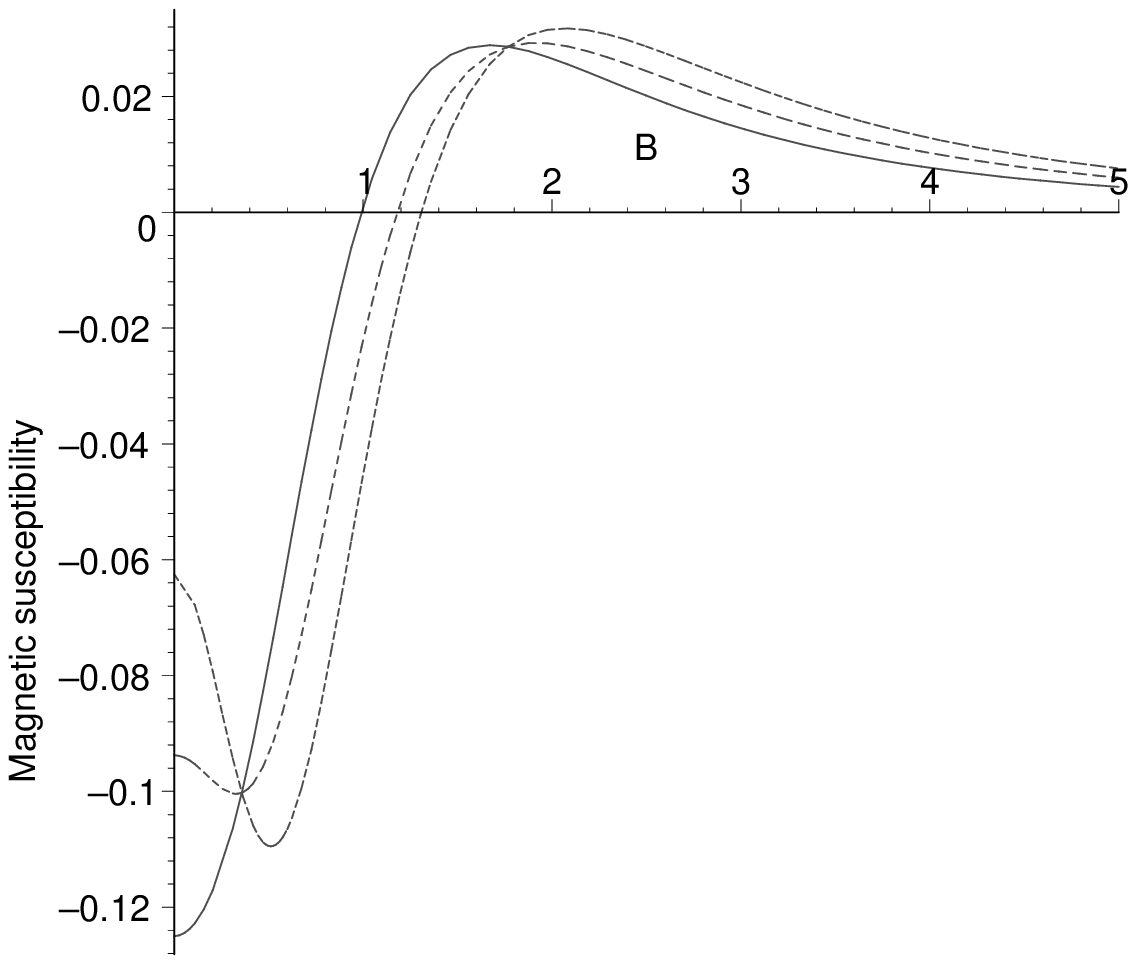}
\end{center}

Figure 1: {\small Magnetic susceptibility vs magnetic field for }$\beta =0$%
{\small \ }$\left( \text{{\small solid line}}\right) ,${\small \ }$\beta
=0.5 ${\small \ }$\left( \text{{\small doted line}}\right) ${\small \ and }$%
\beta =1${\small \ }$\left( \text{{\small dashed line}}\right) .$

\bigskip

Let us examine in details the behavior of the susceptibility in the
following limiting cases:

\begin{itemize}
\item \underline{\textit{Weak magnetic field \ }$\omega <\omega _{0}$\textit{%
:}}

Deriving $\left( \ref{m}\right) $ with respect to the magnetic field, the
susceptibility is given by
\end{itemize}

\begin{equation}
\chi _{\beta }=-\frac{V\mu _{\text{B}}^{2}}{\tilde{\beta}\hbar ^{3}\lambda
^{3}\omega _{0}^{3}}\left[ 1-\frac{m\beta }{\tilde{\beta}}\left( 1-9\frac{%
\omega ^{2}}{\omega _{0}^{2}}\right) \right] ,
\end{equation}%
which shows that the system exhibits the usual Landau diamagnetism in the
regime of high temperatures since we have always $1>\frac{m\beta }{\tilde{%
\beta}}$ by virtue of the GUP. Here we note that we have the value of $B_{1}$
given by $\omega =\omega _{0}/3.$ For zero magnetic field we have

\begin{equation}
\chi _{\beta }=-\frac{V\mu _{\text{B}}^{2}}{\tilde{\beta}\hbar ^{3}\lambda
^{3}\omega _{0}^{3}}\left[ 1-\frac{m\beta }{\tilde{\beta}}\right] .
\label{s}
\end{equation}%
Switching off the minimal length we obtain

\begin{equation}
\chi _{\beta }=-\frac{V\mu _{\text{B}}^{2}}{\tilde{\beta}\hbar ^{3}\lambda
^{3}\omega _{0}^{3}}.  \label{r3}
\end{equation}%
Let us mention that the susceptibility obtained in \cite{Jellal} in the
context of $\theta -$non-commutative quantum mechanics shows an unusual
behavior since it vanishes for $\theta =0$.

Notice that the susceptibility given by $\left( \ref{s}\right) $ is weaker
than the one in the ordinary case since the contribution of the minimal
length is of paramagnetic nature. \ This suggests that the perturbation of
the space by the minimal length generates magnetic moments  in the direction
of the applied external magnetic field. We note also that the limit $%
T\rightarrow \infty $ is forbidden by the condition  $\left( \ref{cond2}%
\right) ,$ and then the susceptibility is always finite. However, we note
that this result is a consequence of the physical statement that, all
physical lengths must be larger than the minimal length.

On the other we obtain $\chi _{\beta }=0$ for a minimal thermal wavelength
given by

\begin{equation}
\lambda _{\text{min}}=\sqrt{\pi }\left( \Delta X\right) _{\text{min}}
\end{equation}%
which in turns define a maximal temperature

\begin{equation}
kT_{\text{max}}=\frac{2}{mc^{2}}\left( \frac{\hbar c}{\left( \Delta X\right)
_{\text{min}}}\right) ^{2}.
\end{equation}

The existence of a maximal temperature have been recently revealed in the
context of the thermodynamics of black holes in the framework of canonical
non-commutative theories \cite{gruppuzo} and with generalized uncertainty
principle \cite{khireddine}. It seems that such a finding is a common
feature of quantum theories on quantized spacetimes.

\begin{itemize}
\item \bigskip \underline{\textit{Strong magnetic field} $\ \omega >>\omega
_{0}:$}
\end{itemize}

In this case the susceptibility is given by

\begin{equation}
\chi _{\beta }=\frac{16V\mu _{\text{B}}^{2}}{3\tilde{\beta}\hbar ^{3}\lambda
^{3}\omega ^{3}}\left[ 1+\frac{5}{3}\frac{m\beta }{\tilde{\beta}}\right] .
\label{strong}
\end{equation}

Here we observe that we have, at high temperatures, orbital paramagnetism
and that $\chi _{\beta }\neq 0$ for finite magnetic field and minimal length.

\section{Conclusion}

In this paper we have investigated the electron magnetism on space where the
coordinate and momentum operators obey generalized commutation relations.
Using the momentum space representation, the eigenstates and the
corresponding energy eigenvalues has been exactly calculated. In the
limiting cases of weak and strong magnetic fields the usual degeneracy of
the Landau levels is now removed by the minimal length. We have also
investigated the magnetic behavior of the system at high temperatures. For
strong magnetic field the contribution of the confining potential is
negligible and we obtain a paramagnetic behavior of the Landau system. The
latter result show, a tendency of the magnetic moments, generated by the
minimal length, to be aligned in the direction of the applied external field
giving then a paramagnetic contribution. For weak magnetic field the orbital
diamagnetism is more pronounced in the standard situation without the
minimal length. For intermediary values of the magnetic field the situation
is inverted and the diamagnetic and paramagnetic behaviors are respectively
stronger and weaker for increasing values of the minimal length. The main
consequence of the minimal length is the existence of a maximal temperature
which renders the susceptibility, in terms of the minimal length, finite.
This important result reflects the regularizing effect of the minimal length.

\section*{Appendix: Radial momentum wave functions}

\renewcommand{\theequation}{A.\arabic{equation}}

\setcounter{section}{0} \setcounter{equation}{0} In this appendix we
calculate, with some details, the radial momentum wave functions.

To solve the eigenvalues equation $H\Psi _{nl }\left( \mathbf{p}\right)
=E_{nl}\Psi _{nl }\left( \mathbf{p}\right) $ with $H$ given by $\left( \ref%
{H2}\right) ,$ in the momentum space representation, we exploit the
rotational invariance of the problem and write $\Psi _{nl }\left( \mathbf{p}%
\right) $ as
\begin{equation}
\Psi _{nl }\left( \mathbf{p}\right) =\frac{e^{\frac{i}{\hbar }l\phi }}{\sqrt{%
2\pi \hbar }}R_{nl}(p),
\end{equation}%
where $n$ is the radial quantum number and $l$ \ the magnetic number.

Using the two dimensional representation of the position operators given by
eq.(\ref{x-p}) we have the following differential equation for the radial
part of the wave function

\begin{eqnarray}
&&\left( \left( 1+\beta p^{2}\right) \frac{\partial }{\partial p}\right)
^{2}R_{nl}(p)+\left( 1+\beta p^{2}\right) ^{2}\left( \frac{1}{p}\frac{%
\partial }{\partial p}-\frac{l^{2}}{p^{2}}\right) R_{nl}\left( p\right)
\notag \\
&&{-}\left( \frac{\omega l}{2m\tilde{\omega}^{2}\hbar }+\frac{p^{2}}{\left(
m\hbar \tilde{\omega}\right) ^{2}}+\mathcal{E}_{nl}\right) R_{nl}\left(
p\right) {=0}  \label{eq0}
\end{eqnarray}%
with $\mathcal{E}_{nl}$ given by
\begin{equation}
\mathcal{E}_{nl}=\frac{2E_{nl}}{m\hbar ^{2}\tilde{\omega}^{2}}-\frac{%
p_{z}^{2}}{\left( m\hbar \tilde{\omega}\right) ^{2}}.
\end{equation}%
In terms of the new variable $\xi =\frac{1}{\sqrt{\beta }}$arctan$\left( p%
\sqrt{\beta }\right) $ we write $\left( \ref{eq0}\right) $ as
\begin{eqnarray}
&&R_{nl}^{\prime \prime }(\xi )+\sqrt{\beta }\left( \text{cot}\sqrt{\beta }%
\xi +\text{tan}\sqrt{\beta }\xi \right) R_{nl}^{\prime }(\xi )-\beta
l^{2}\left( \text{cot}\sqrt{\beta }\xi +\text{tan}\sqrt{\beta }\xi \right)
^{2}R_{nl}(\xi )  \notag \\
&-&\left( \Omega +\frac{1}{\beta (m\tilde{\omega}\hbar )^{2}}\right) \text{%
tan}^{2}\sqrt{\beta \rho }R_{nl}(\xi )+\left( \mathcal{E}_{nl}-{\Omega }%
\right) R_{nl}(\xi )=0,  \label{eq2}
\end{eqnarray}%
with $\Omega =\frac{l\omega }{2m\tilde{\omega}^{2}\hbar }$. We simplify $%
\left( \ref{eq2}\right) $ by setting $R_{nl}(\xi )=c^{\lambda }f(s)$ with $c$
and $s$ defined as
\begin{equation}
c=\text{cos}\sqrt{\beta }\xi ,\qquad s=\text{sin}\sqrt{\beta }\xi .
\end{equation}%
A straightforward calculation gives the following differential equation for $%
f(s)$
\begin{eqnarray}
&&\left( 1-s^{2}\right) f^{\prime \prime }(s)+\left( \frac{1}{s}-\left(
2\lambda +1\right) \right) f^{\prime }(s)  \notag \\
&+&\left( \left( \lambda \left( \lambda -2\right) -\frac{1}{\kappa ^{4}}%
-l^{2}\right) \frac{s^{2}}{c^{2}}+\left( \frac{\mathcal{E}_{nl}-\Omega }{%
\beta }-2\lambda -l^{2}\right) -\frac{l^{2}}{s^{2}}\right) f(s)=0,  \label{f}
\end{eqnarray}%
where we have set $\kappa =\sqrt{m\tilde{\omega}\hbar \beta }$. Then we
cancel the term with $\frac{s^{2}}{c^{2}}$ by choosing $\lambda $ such that
\begin{equation}
\lambda ^{2}-2\lambda -l^{2}-\frac{1}{\kappa ^{4}}=0.
\end{equation}%
The solutions of this equation are given by
\begin{equation}
\lambda =1\pm \frac{1}{m\tilde{\omega}\hbar \beta }\sqrt{1+\left( m\hbar
\tilde{\omega}\beta \right) ^{2}\left( 1+l^{2}\right) }  \label{lam}
\end{equation}%
The next step is to cancel the centrifugal barrier in equation (\ref{f}) by
setting $f(s)=s^{\left\vert l\right\vert }g(s)$. Then we have
\begin{equation}
(1-s^{2})g^{\prime \prime }(s)+\left( \frac{2\left\vert l\right\vert +1}{s}%
-(2\lambda +2\left\vert l\right\vert +1)s\right) g^{\prime }(s)+\left( \frac{%
\mathcal{E}_{nl}-\Omega }{\beta }-2l^{2}-2\lambda (\left\vert l\right\vert
+1)\right) g(s)=0.
\end{equation}%
At this stage we use the variable $z=2s^{2}-1$ to obtain
\begin{equation}
(1-z^{2})g^{\prime \prime }(z)+\left[ (\left\vert l\right\vert -\lambda
+1)-(\left\vert l\right\vert +\lambda +1)z\right] g^{\prime }(z)+\frac{1}{4}%
\left( \frac{\mathcal{E}_{nl}-\Omega }{\beta }-2l^{2}-2\lambda (\left\vert
l\right\vert +1)\right) g(z)=0  \label{eq}
\end{equation}%
Defining
\begin{equation}
a=\lambda -1,\qquad b=\left\vert l\right\vert ,
\end{equation}%
and imposing the following condition, to get a polynomial solution,
\begin{equation}
\frac{\mathcal{E}_{nl}-\Omega }{\beta }-2L^{2}-2\lambda (\left\vert
l\right\vert +1)=4n(n+a+b+1),  \label{cond0}
\end{equation}%
with $n$ a non negative integer, we reduce $\left( \ref{eq}\right) $ to the
following form
\begin{equation}
(1-z^{2})g^{\prime \prime }(z)+\left[ (b-a)-(a+b+2)z\right] g^{\prime
}(z)+n(n+a+b+1)g(z)=0.  \label{eq1}
\end{equation}%
The solutions of equation (\ref{eq1}) are given by Jacobi polynomials
\begin{equation}
g(z)=P_{n}^{(a,b)}(z).
\end{equation}%
Using the old variable $p$, the radial part of the wave function is then
given by
\begin{equation}
R_{nl}(p)={\mathcal{N}}(1+\beta p^{2})^{-\frac{\lambda +\left\vert
l\right\vert }{2}}(\beta p^{2})^{\frac{\left\vert l\right\vert }{2}%
}P_{n}^{(\lambda -1,\left\vert l\right\vert )}\left( \frac{\beta p^{2}-1}{%
\beta p^{2}+1}\right) ,
\end{equation}%
where ${\mathcal{N}}$ is a normalization constant.

\providecommand{\href}[2]{#2}\begingroup\raggedright

\end{document}